\documentstyle[prl,aps,twocolumn,epsf,floats]{revtex}
\begin{document}
\draft
\preprint{}
\title{Laser induced freezing of charge stabilized colloidal system}
\author{Chinmay Das $^1$, A. K. Sood $^2$ \cite{d} and
  H. R. Krishnamurthy $^3$ \cite{d}}
\address{Department of Physics, Indian Institute of Science,
Bangalore 560 012, India\\
$^1$ e-mail : cdas@physics.iisc.ernet.in
$^2$ e-mail : asood@physics.iisc.ernet.in
$^3$ e-mail : hrkrish@physics.iisc.ernet.in}
\date{\today}
\maketitle

\begin{abstract}
We present results from an extensive simulational study of the modulated
liquid $\longleftrightarrow$ crystal transition in a 2-d 
charge-stabilized colloid subject to a 1-d
laser field modulation commensurate with the crystalline phase.  
{\em {Contrary to some earlier simulational
and experimental findings we do not find any reentrant liquid phase
in our simulation. 
Furthermore the transition remains first-order (albeit weak, with a fairly large
correlation length) even in the limit of infinite field, contrary to 
mean-field predictions}}.
In the modulated liquid 
phase,
while the translational order decays exponentially, 
the bond orientational order is actually long ranged.

\end{abstract}

The freezing
of charge-stabilized colloidal particles in the presence of a
stationary modulation (laser) field has been of considerable research 
interest in recent years. Using light scattering techniques
Chowdhury, Ackerson and Clark \cite{expt:ackerson}
showed that a charge stabilized colloidal liquid system, confined between
two glass plates to form a single layer, when subjected to 
a commensurate stationary laser modulation (i.e.\ with the wavevector of
the modulating laser field tuned to be half the  wavevector at which the 
direct correlation function $c^{(2)}(q)$ of the liquid develops its 
first peak in the absence of an external potential), 
freezes to form a triangular
lattice with full two dimensional symmetry.
They \cite{expt:ackerson}
also analyzed the phenomenon in terms of a simple 
Landau - Alexander - McTague (LAM)
\cite{th:landau} theory and concluded that the transition from the
modulated liquid to the crystal becomes continuous (second-order) when
the field strength is large.
The continuous growth of intensities produced by all the density
modes with increasing external field intensity was taken by 
Ackerson and Chowdhury \cite{expt:ackerson_faradey} as an indication 
in favor of a second-order transition scenario.
 Later experimental studies involving direct
microscopic observations \cite{expt:Loudiyi} and simulational studies
using Monte-Carlo (MC) technique \cite{sim:Loudiyi} confirmed this 
phenomenon of laser induced freezing (LIF). 

Several authors \cite{th:DFT:Xu}, \cite{th:DFT:Barrat}, \cite{th:DFT:jay}
have studied this problem using
density functional theory \cite{th:DFT:RY}. The authors of
\cite{th:DFT:jay} have shown from general symmetry grounds that 
for a suitably chosen modulation potential the
free energy expansion  for the crystalline phase about the modulated 
liquid contains the relevant order parameters only in even powers. Hence 
there arises a possibility of change over of the first-order 
freezing transition at
low external field strengths to a continuous transition for large enough
external field via a tricritical point. Their DFT results were in 
accordance with these symmetry arguments.

In a later work Chakrabarti {\it et al.} \cite{sim:MC:jay} studied a 
two dimensional colloidal system using MC simulation. 
The use of a standard procedure of looking at
finite-size behavior of the
fourth cumulant of the energy \cite{th:vl:all} seemed to confirm  
the existence of the
tricritical point. Their study also found
an intriguing reentrant modulated liquid phase - where increasing the 
external field strength actually melted the
system instead of taking it towards the crystalline phase. Another unusual
feature found in their simulation was that 
the order parameter jumped to zero at a lower value
of the screening length compared to where the specific heat peak was seen.

Very recently Wei {\it et al.} \cite{expt:konstanz} reported an 
experimental study of LIF and from the decay of pair correlation
and the real space density plots concluded that their
results were in accordance with the results of reference 
\cite{sim:MC:jay}, showing a reentrant liquid phase.

In this letter we report the main results obtained from a vastly more 
extensive Monte-Carlo simulation than that in ref. \cite{sim:MC:jay}
of charge stabilized colloidal particles (with diameter 
$2 R = 1.07 \mu m$). They are assumed to be confined in a 2-d cell of size 
$ \frac{\sqrt{3}}{2} a_s L \times a_s L $ with periodic boundary conditions
and subjected to an external potential of the form
$U(\vec{r}) = - {\cal V}_e cos(q_0 x)$, with $q_0 = 2 \pi /
(\frac{\sqrt{3}}{2} a_s)$,
where $a_s$ is the mean inter-particle separation. The inter-particle 
interaction is modeled by the DLVO potential:
\begin{equation}
U_{ij}(r) = \frac{(Z e)^2}{\epsilon} 
(\frac{exp(\kappa R)}{1 + \kappa R})^2
\frac{exp(-\kappa r_{ij})}{|r_{ij}|}
\label{eq:DLVO}
\end{equation}
Here $Z e$ ($Z = 7800$) is the effective surface charge,
$\epsilon$ (=78) is the dielectric constant of the solvent and
$\kappa$ is the inverse of the Debye screening length due to
the counterions in the solvent. Details of the simulation will be
published elsewhere \cite{yet_to_be}.
\begin{figure}[htbp]
\epsfxsize=6cm
\epsfysize=4.5cm
\centerline{\epsfbox{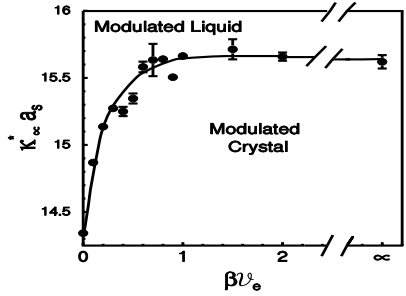}}
\caption{Phase diagram of laser induced freezing (extrapolated 
for infinite system size).}
\label{fig:phdiag}
\end{figure}

\begin{figure}[htbp]
\epsfxsize=7.5cm
\epsfysize=8cm
\centerline{\epsfbox{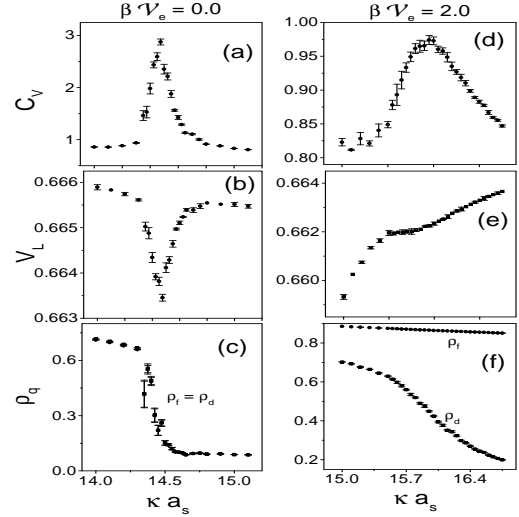}}
\caption{Simulation results for 400 particles. The left hand panels 
(a-c) are for $\beta {\cal V}_e = 0.0$ and the right hand 
panels (d-f) are for $\beta {\cal V}_e = 2.0$.}
\label{fig:simmeasure}
\end{figure}

The phase diagram obtained by extrapolating the simulation results 
to infinite system size as detailed below
is shown in figure \ref{fig:phdiag}.  {\em {It does not have the
reentrance reported in } }\cite{sim:MC:jay}.
The transition is ``strongly first-order'' at zero field, signaled 
by a strong peak in the specific heat $C_V$ 
[figure \ref{fig:simmeasure}(a)] and
sharp rise of the order parameters \cite{th:DFT:jay} ,\cite{sim:MC:jay}
[figure \ref{fig:simmeasure}(c)], at 
$\kappa^*_L a_s = 14.47$.  At larger field strengths, 
the peak in $C_V$ becomes less prominent [figure \ref{fig:simmeasure}(d)]
and the order parameter $\rho_d$ characterizing the modulated liquid 
$\longleftrightarrow$ crystal transition \cite{th:DFT:jay}, 
\cite{sim:MC:jay}  seems to go to zero
continuously [figure \ref{fig:simmeasure}(f)], in apparent accordance with
the mean field theory predictions \cite{expt:ackerson} , 
\cite{th:DFT:jay}. 
{\em { However, as we show below, in actuality  the transition 
seems to remain first-order (albeit weak, with a fairly large
correlation length) even in the limit of infinite field, contrary to 
the mean-field predictions}}.

For a finite system, one expects that the transition will actually 
be rounded always. Hence one must
look at the scaling of various quantities with large system size $L$
to ascertain the order of the transition. 
In particular the limiting value  of the fourth cumulant of energy
$V_L \equiv 1 - \frac{<E^4>_L}{3 <E^2>^2_L}$ and its scaling 
behavior have been widely used  to judge the order
of the transition \cite{th:vl:all}, \cite{sim:MC:jay}. In case of a 
first-order transition, $V_{\infty} = 2/3$,
the peak in $C_V$ should diverge as $L^d$, while 
$(\kappa_L^* - \kappa_{\infty}^*)$ and $(V_L - V_{\infty})$ 
should scale as $L^{-d}$.
For small system sizes, there are corrections to this scaling behavior 
and we show below that at large field strengths these corrections
become even more important.

For a first-order transition scenario, assuming that the probability
distribution for energy (per particle) is the sum of two Gaussian  
distributions centered around $E_+$ and $E_-$
corresponding to the disordered (modulated liquid) and the ordered 
(crystalline) phases, one can derive the following results,
\cite{th:vl:all}, \cite{yet_to_be}: 
\begin{eqnarray}
C_L^*=A L^d + B + D L^{-d} + {\cal O}(L^{-2 d}), \label{eq:cvpeak}\\
\kappa_L^*  = \kappa_{\infty}^* + M L^{-d} + N L^{-2d} + 
{\cal O}(L^{-3 d}), \label{eq:cvshift}
\end{eqnarray}
where $C_L^*$ refers to the height of the specific heat peak for 
a system of linear size $L$ and $A = \frac{(E_+ - E_-)^2}{4 k_B T^2}$.
To leading order, the constants
$M, N$ and $D$ are proportional to the inverse of 
$\frac{\partial{\Delta F}}
{\partial{\kappa}}$, $\Delta F$ being the free energy difference 
between the liquid and the crystalline phases.

\begin{figure}[htbp]
\epsfxsize=7.5cm
\epsfysize=6.5cm
\centerline{\epsfbox{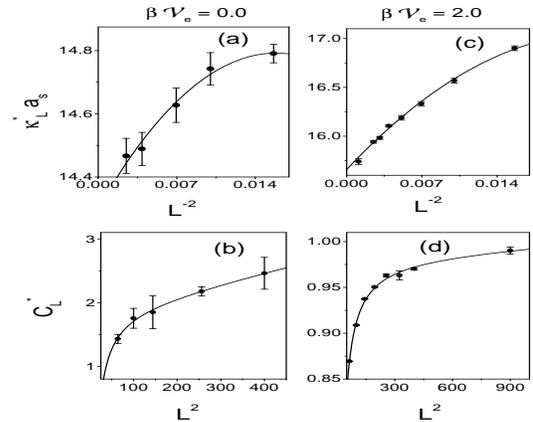}}
\caption{Finite size scaling behavior of the transition 
$\kappa_c(L)$ (a,b) and of the specific heat peak height (c,d)}
\label{fig:scalfnt}
\end{figure}

Figures \ref{fig:scalfnt}(a) and (c) show the scaling behavior of 
$\kappa_L^*$  (from the specific heat peak) for $\beta {\cal V}_e = 0.0$ 
and $\beta {\cal V}_e = 2.0$ respectively. 
By fitting equation \ref{eq:cvshift} (the solid line),  we get 
$\kappa_{\infty}^* a_s \cong 14.34 \pm 0.02 (15.66 \pm 0.03)$ for 
$\beta {\cal V}_e = 0.0 (2.0)$.
$C^*_L$ is plotted as function of $L$ in
figure \ref{fig:scalfnt}(b) and (d). By fitting equation
\ref{eq:cvpeak} we find $\beta (E_+ - E_-) \cong 0.071 \pm 0.0067 
(0.0044 \pm 0.0031)$ for $\beta {\cal V}_e = 0.0 (2.0)$.

While the above analysis indicates that a first-order scenario is a 
likely possibility even at large $\beta {\cal V}_e$, the 
scaling analysis is done for results which span
less than a decade in $L$ ($L = 8$ to $30$). Because the peak in $C_V$ is 
rather broad and small in height, one needs to average over several 
million MC steps to determine the
peak position and peak height within acceptable errorbars. This limits 
the largest system we are able to simulate to $L = 30 $.

One can further argue that even though the transition remains first 
order and does not show any reentrant melting till 
$\beta {\cal V}_e = 2.0$, there is always
a possibility that the transition does become continuous at a  
larger field strength. To settle this we have carried out an 
even more extensive simulation at $\beta {\cal V}_e = \infty$, so that 
the particles are confined to parallel lines. The resulting 
simplifications allow us to write a more efficient MC code, 
making simulations upto $L = 100$ doable within our computing resources.
\begin{figure}[htbp]
\epsfxsize=7.5cm
\epsfysize=6.5cm
\centerline{\epsfbox{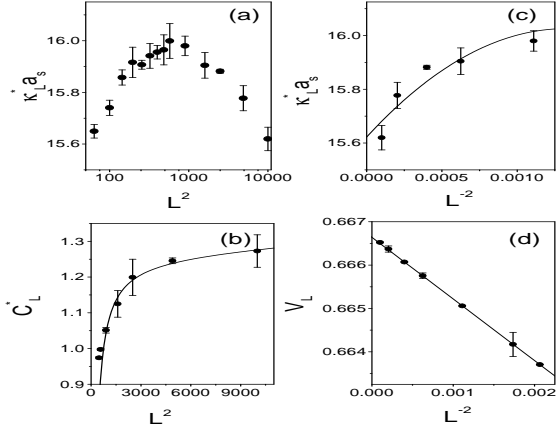}}
\caption{Finite size scaling for $\beta {\cal V}_e = \infty$}
\label{fig:infcv}
\end{figure}
The plot of
$\kappa_L^*$ as a function of $L^2$ for $\beta {\cal V}_e = \infty$, 
shown in  Figure \ref{fig:infcv}(a), has a maximum around $L \sim 30 a_s$.
This non-monotonic behavior seem to suggest a large but finite length 
scale in the system. The existence of such a length scale is 
difficult to understand in the frame work of a second-order 
transition scenario, where, asymptotically close to the phase transition,
one should have just two length scales, the system size $L$ (to which  
the ``diverging'' correlation length $\xi$ saturates) and the 
interaction scale (which is not much larger than $a_s$ for the present 
case). But if the transition is weakly first-order with a large but 
finite $\xi$, one expects the finite-size scaling behavior of the 
first-order transition discussed above to be valid
only for $L \gg \xi$.
Figures \ref{fig:infcv}(b) and \ref{fig:infcv}(c) show that 
$\kappa_L^*$ and $C_L^*$ for large $L$ do follow a first-order like
scaling [cf., eq.s (2) and (3)], 
with $\kappa_{\infty}^* \cong 15.62 \pm 0.05$ and 
$\beta(E_+ - E_-) \cong 0.0014 \pm 0.0004$; 
so does $V_L$ which scales
as $L^{-2}$ as shown in figure \ref{fig:infcv}(d). 
\begin{figure}[htbp]
\epsfxsize=7.5cm
\epsfysize=6.5cm
\centerline{\epsfbox{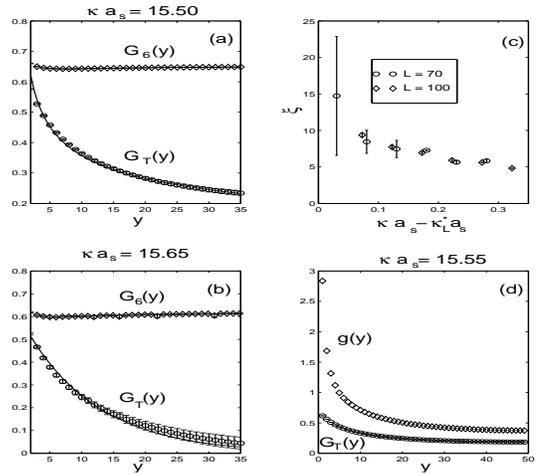}}
\caption{correlation functions for $\beta {\cal V}_e = \infty$}
\label{fig:corfn}
\end{figure}

For $\beta {\cal V}_e = \infty$ we have computed the correlation 
functions for the coarse grained translational and
bond-orientational order parameters, to be denoted by $G_T(y)$ and  
$G_6(y)$ respectively, for $L = 70$ and $100$.
In the modulated liquid phase we obtain 
$\xi$ by fitting $G_T(y)$ to a decaying exponential.
Figure  \ref{fig:corfn}(c) 
shows that $\xi$ actually becomes large even
in the liquid phase as one goes towards  $\kappa_L^* $. But it
saturates at around $18 a_s$, indicating that the transition
really is a weakly first-order transition.

Figures \ref{fig:corfn}(a) and (b) show $G_6(y)$
for the liquid and the solid phase near $\kappa_L^*$. 
As can be seen from the figures, 
$G_6(y)$ remains long ranged even in the liquid phase. A 2-d liquid
has a local hexagonal order. When one applies a commensurate field, a 
unique direction (modulo $60^o$) is picked up every where in the system. 
Thus the 1-d modulation potential induces not only translational 
ordering, but also
a long-range orientational order even in the modulated liquid phase. 
This is also understandable from the general phenomenological 
mean-field free energy given in \cite{nelson}, 
where  a term linear in the bond-orientational order parameter 
arises as soon the translational order parameter $\rho_f$ 
conjugate to ${\cal V}_e$ is turned on in the modulated liquid phase.
Thus in the modulated liquid, $G_6(y)$ goes to a
constant value at large distances. What distinguishes
the modulated liquid from the crystalline phase is 
$G_T(y)$, which decays exponentially in the modulated liquid 
phase but as a power law in the 2-d crystalline  phase. The
energy difference between the two phases becomes smaller as 
$\beta {\cal V}_e$ is increased (though it
never becomes strictly zero) and is responsible for the
broad specific heat peak and the broad region over which the order 
parameter goes to zero.
We believe that this probably led to the results reported in 
\cite{sim:MC:jay}, where the true transition was missed, and a 
statistical fluctuation was mistakenly taken to be the 
$C_V$ peak and hence the signature of a transition.
Since different moments of energy are not independent,
a chance increase of $C_V$ will be associated with a 
dip of $V_L$. Such artifacts arising from insufficient averaging are 
presumably absent in the present simulation due to the 
extensive averaging that has been carried out. This also
explains why the earlier simulation \cite{sim:MC:jay} failed to see 
the order parameter fall to zero until much
higher values of $\kappa a_s$.

\begin{figure}
\epsfxsize=8cm
\epsfysize=6.5cm
\centerline{\epsfbox{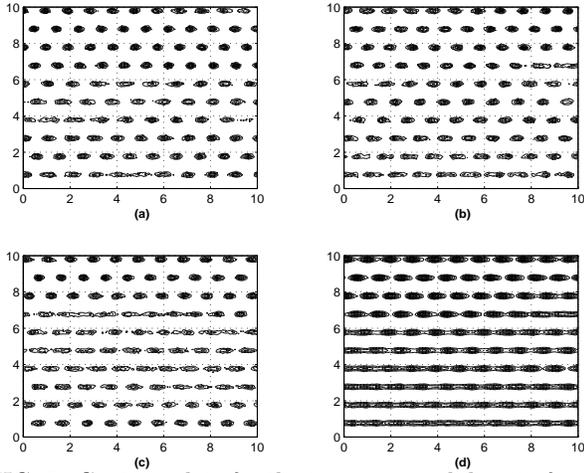}}
\caption{Contour plots for the time averaged density from a 
Brownian dynamics simulation at
$\beta {\cal V}_e = 2.0$ and $\kappa a_s = 15.3$. The 400 particle 
system was equilibrated in a run of 20 minutes. Figures (a), (b) 
and (c) correspond to averaging over 2 minute 
intervals starting from the 0 th, 6 th and 16 th minute
respectively after the system had equilibrated. Figure (d) corresponds
to averaging over 20 minutes starting from equilibration.}
\label{fig:bdavden}
\end{figure}

For moderate external fields, substantial diffusion precedes the actual 
transition. While maintaining the translational order over the full 
system, locally the system accommodates defects. In Figure 
\ref{fig:bdavden} we show contour plots for the time averaged density as 
computed over different time spans from a Brownian dynamics simulation 
of 400 particles for $\beta {\cal V}_e = 2.0$. The short-time 
averaged density plots show near perfect crystalline  order, with 
small regions where the particles have moved considerably from the 
lattice positions.  If one averages for a long time ($\simeq 20$ 
minutes of real time), the average density
shows that the chance of finding a particle is nonzero everywhere 
along the lines. But we still find that the peaks of the average 
density maintain
the crystalline order. Scattering experiments, which measure the 
structure at short time scales compared to the diffusion time, 
can differentiate this from the liquid
phase via the existence of Bragg peaks. And in our simulation we can 
directly compute the order parameter,
which, for the above parameter values, is as large as $0.75$! 

The claim in \cite{expt:konstanz} of the observation of reentrant
melting into the modulated liquid phase at high $\beta {\cal V}_e $
was based on the observed stronger decay of the envelope of $g(r)$ than
that theoretically predicted for the crystalline state \cite{nelson}. 
The problem with this analysis is that the correct power-law decay 
in the envelope of  $g(r)$ sets in at much larger distances than for
the order parameter correlation $G_T(y)$ (which unfortunately is
difficult to measure from experiments).  In fig.\ \ref{fig:corfn}(d)
we plot the envelope of $g(y)$ and $G_T(y)$ for 
$\beta {\cal V}_e = \infty$, and $\kappa a_s =15.55$ , i.e., in
the crystalline phase but close to the transition . 
The envelope of $g(y)$ falls much more rapidly than
$G_T(y)$ even upto $y = 20$ , and would yield a large exponent if 
fitted to a power law, in contrast to $G_T(r)$  which quickly
attains a power law decay with exponent $-0.3$. We expect that 
for the same parameter values where \cite{expt:konstanz} suggested 
a modulated liquid phase, light scattering experiments or 
order parameters calculated from Fourier transforming individual 
configurations prior to averaging will 
reveal the existence of a crystalline order.

In conclusion we have presented new simulational evidence 
that the laser induced freezing in charge stabilized colloids 
remains first-order for an arbitrarily large laser field, 
{\em {in contrast to all theoretical predictions hitherto} }.
The reentrant liquid phase suggested in \cite{sim:MC:jay}
and  \cite{expt:konstanz} is absent in the  resulting  phase
diagram. Also we have shown that the modulated liquid phase 
has long range bond orientational order. 
The difference in the nature of transition between that predicted 
by the DFT  and that found in our  simulation is likely to be 
due to fluctuations, which are neglected in the DFT calculations 
in [6-8]; the inclusion of such fluctuations 
is known to turn some transitions first-order,
eventhough they are predicted to be continuous by 
mean-field theory \cite{fluctuations} \cite{binder}.

\section*{Acknowledgments}
\label{sec.ack}
The authors thank T.V. Ramakrishnan,  S. Sengupta, S.S. Ghosh, 
J. Chakrabarti, R. Pandit,
C. DasGupta and S. Ramaswamy for many useful discussions. 
We thank SERC, IISc for 
computing resources. CD thanks CSIR, India for
financial support.

\end{document}